\pdfoutput=1

\documentclass{article}

\usepackage[numbers]{natbib}

\usepackage[preprint]{neurips_2023}

\usepackage[utf8]{inputenc} 
\usepackage[T1]{fontenc}    
\usepackage{hyperref}       
\usepackage{url}            
\usepackage{booktabs}       
\usepackage{amsfonts}       
\usepackage{nicefrac}       
\usepackage{microtype}      
\usepackage{xcolor}         
\usepackage{amsmath}
\usepackage[pdftex]{graphicx}
\usepackage{multirow}
\usepackage[all]{hypcap}

\title{Elixir: Train a Large Language Model on a Small GPU Cluster}

\author{%
  Haichen Huang\\
  HPC-AI Technology Inc. \\
  \texttt{hhc@hpcaitech.com} \\
  \And 
  Jiarui Fang \thanks{This work was done when Jiarui worked at HPC-AI Technology Inc.} \\
  HPC-AI Technology Inc. \\
  \texttt{fangjr@hpcaitech.com} \\
  \And 
  Hongxin Liu \\
  HPC-AI Technology Inc. \\
  \texttt{liuhongxin@hpcaitech.com} \\
  \AND 
  Shenggui Li \\
  HPC-AI Technology Inc. \\
  \texttt{lisg@hpcaitech.com} \\
  \And 
  Yang You \thanks{Dr. You is a faculty member at NUS. This work was done at HPC-AI Technology Inc.}\\
  National University of Singapore \\
  \texttt{youy@comp.nus.edu.sg} \\
}

\begin{document}

\maketitle

\begin{abstract}
  In recent years, large language models have achieved great success due to their unprecedented size.
  However, training these models poses a challenge for most researchers as it requires a substantial number of GPUs.
  To reduce GPU memory usage, memory partitioning and memory offloading have been proposed.
  These approaches eliminate memory redundancies and offload memory usage to the CPU and NVMe memory, respectively, enabling training on small GPU clusters.
  However, directly deploying these solutions often leads to suboptimal efficiency.
  Only experienced experts can unleash the full potential of hardware by carefully tuning the distributed configuration.
  Thus, we present a novel solution, Elixir, which automates efficient large model training based on pre-runtime model profiling.
  Elixir aims to identify the optimal combination of partitioning and offloading techniques to maximize training throughput. 
  In our experiments, Elixir significantly outperforms the current state-of-the-art baseline. 
  Our optimal configuration achieves up to a 3.4$\times$ speedup on GPT-2 models compared with SOTA solutions.
  We hope that our work will benefit individuals who lack computing resources and expertise, granting them access to large models\footnote[1]{The beta version of Elixir is now available at \url{https://github.com/hpcaitech/ColossalAI/tree/feature/elixir}}.
  
\end{abstract}

\section{Introduction}

The current success of deep learning (DL) is attributed to the rise of pre-trained large language models (LLMs) \cite{bert-devlin2018bert, opt-zhang2022opt, gpt2-radford2019language, gpt3-brown2020language, gpt4-OpenAI2023GPT4TR, bloom-scao2022bloom, palm-chowdhery2022palm, lamda-thoppilan2022lamda}.
LLMs are widely used not only in NLP applications such as conversation, Q\&A, and text generation, but also in multimodal tasks such as image generation \cite{parti-yu2022scaling, muse-chang2023muse}, and speech synthesis \cite{speech-wang2023neural}.
However, training LLMs remains challenging due to the growing size of models and limited GPU memory.
In the past five years, the largest dense models have significantly increased in size, from 340 million parameters in BERT \cite{bert-devlin2018bert} to 540 billion parameters in PaLM \cite{palm-chowdhery2022palm}. 
Exerting the power of the mixture-of-experts architecture \cite{moe-shazeer2017outrageously}, the number of parameters in the largest sparse model \cite{switch-fedus2022switch} has exceeded 1 trillion.
Meanwhile, GPU memory has only increased to 80GB \cite{nvidia-A100, nvidia-H100}.

To address the memory bottleneck, researchers have proposed distributed training techniques. 
Distributed data parallelism (DDP) divides input data and assigns each device to compute its partition simultaneously.
Though DDP accelerates training, it requires each device to store a complete model copy.
Zero Redundancy Optimizer (ZeRO) \cite{zero-3-rajbhandari2020zero} is designed to eliminate memory redundancy in DDP.
ZeRO partitions the model among devices, gathers parameters before computations, and scatters them afterward.
The communication volume in ZeRO scales with the model size.
Tensor parallelism (TP) partitions parameters used in matrix multiplication (MM) operators and replaces the original MM computation with parallel MM algorithms.
Megatron-LM \cite{megatron-shoeybi2019megatron} employs an efficient TP implementation that eliminates parameter redundancy but duplicates a part of intermediate variables during training. 
The communication volume in this approach is determined by the size of intermediate variables, which grows linearly with batch size, sequence length, and model size.
Pipeline parallelism (PP) \cite{gpipe-huang2019gpipe, pipedream-narayanan2019pipedream, 3d-parallel-narayanan2021efficient, chimera-li2021chimera} distributes the layers of the model across compute devices, and treats the devices as a data processing pipeline.
PP generally has the lowest communication costs but it brings waiting bubbles.

Though memory partitioning addresses the memory bottleneck, it requires the aggregate GPU memory to exceed the model size.
Memory offloading, on the other hand, leverages next-level memory space such as CPU memory and NVMe disks to store large models when the aggregate GPU memory is insufficient. 
ZeRO-Infinity \cite{zero-infinity-rajbhandari2021zero} demonstrates that memory offloading is the most promising method for scaling up models since memory and storage hardware are generally more affordable than GPUs.

\begin{figure}[t]
    \centering
    \includegraphics[width=0.9\linewidth]{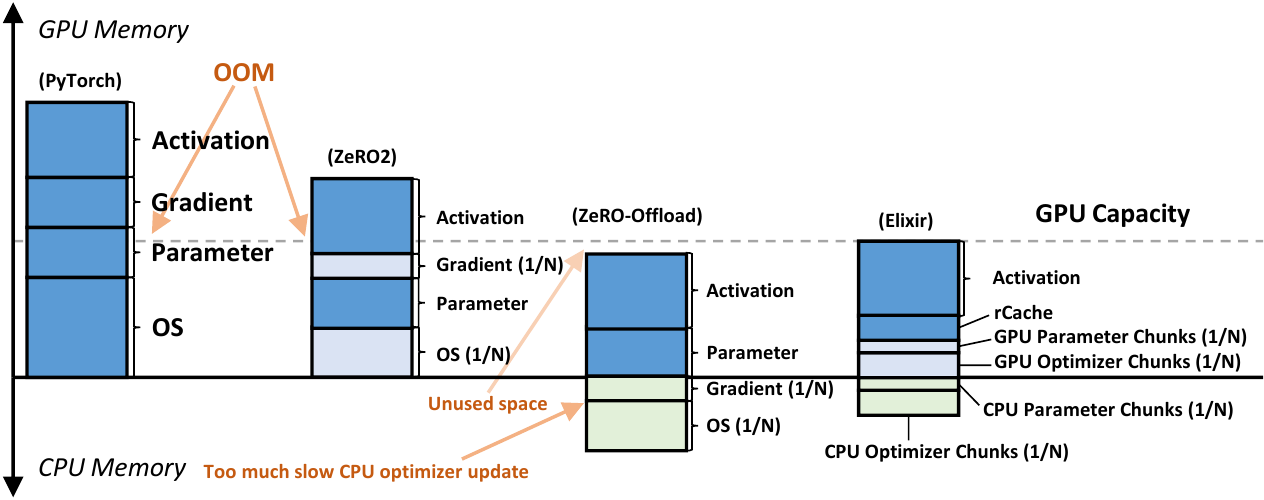}
    \caption{Comparison between Elixir and other distributed solutions.}
    \label{fig:compare}
\end{figure}

ZeRO-Infinity, despite being a powerful heterogeneous training system, has two significant drawbacks: 
(1) ZeRO requires users to configure numerous stages and arguments. 
It is a rather intricate and time-consuming task to explore distributed configurations. 
ZeRO optimization involves 18 adjustable arguments, named \textit{stage3\_max\_live\_parameters}, \textit{sub\_group\_size}, \textit{round\_robin\_gradients}, etc., which might be unfamiliar to most AI researchers.
Although DeepSpeed \cite{ds-rasley2020deepspeed} offers an auto-tuning command, it only considers memory partitioning and does not incorporate offloading.
(2) ZeRO achieves suboptimal efficiency due to a lack of awareness of the training process.
By default, ZeRO-3 partitions all parameters, leading to additional communication for gathering parameters during the forward and backward pass.
Similarly, its offloading option offloads all optimizer states to CPU memory, resulting in slow CPU optimizer updates for all parameters.
We argue that using free GPU memory to store optimizer states or retaining gathered parameters during training can improve training throughput.

In this paper, we propose Elixir, an approach to automate large model training on a small GPU cluster.
Our main goal is to make the best use of all GPU memory, maximizing the training throughput.
To achieve this, we develop a pre-runtime profiler that accurately measures memory usage before training.
Our profiler can profile the 175B OPT \cite{opt-zhang2022opt} model on a single A100 GPU within 10 seconds. 
We compact parameters into continuous blocks of the same length, called chunks, as the memory units in our parallel training system for convenience.
We introduce rCache to achieve fine-grained control over the degree of memory redundancy.
Therefore, we can adjust rCache size or move chunks around to utilize free GPU memory.
Based on the profiling results, we analyze the benefits of these two options and identify an optimal configuration for training.
In our experiments, our optimal configuration runs much faster than current SOTA solutions. 

Taking Fig.\ref{fig:compare} as an example, PyTorch \cite{pytorch-paszke2019pytorch} DDP triggers a CUDA Out-Of-Memory (OOM) error in this case. 
Although ZeRO-2 partitions gradients and optimizer states, it still triggers an OOM error. 
ZeRO-Offload \cite{zero-offload-ren2021zero} resolves the OOM issue, but it suffers from slow CPU optimizer updates and unused CUDA memory. 
However, Elixir overcomes these challenges by effectively utilizing all GPU memory.

In summary, our contributions are as follows: \\
$\bullet$ We build a pre-runtime profiler designed for large models. 
It is capable of obtaining the computation graph and the memory usage of the model before training. 
We bring this powerful tool to support large model profiling. \\
$\bullet$ We introduce rCache to control the degree of memory redundancy.
Moreover, we build a search engine to find the optimal configuration, maximizing training efficiency automatically. 
Different from previous works, our optimal configuration considers both memory partitioning and memory offloading. \\
$\bullet$ We conduct evaluations on a large scale by testing various model sizes, GPU capacities, numbers of GPUs, and batch sizes. 
When compared to current SOTA solutions, we observe that Elixir achieves up to 3.4$\times$ acceleration without manual tuning.

\section{Background: Techniques for Large Model Training}

The key idea of Elixir is to intelligently manage GPU memory.  
We apply established techniques to reduce memory usage during training. 
First, we discuss how memory is consumed during training.

\textbf{Memory Usage}. 
Memory usage during training primarily consists of five components: parameters, gradients, optimizer states, activations, and buffers.
Optimizer states are the extra memory footprint consumed by the optimizer. 
For example, Adam \cite{adam-kingma2014adam} needs to store averaged momentum and variance of gradients.
We refer to parameters, gradients, and optimizer states collectively as model states. 
Activations are the intermediate temporary variables generated during training. 
Typically, activations are stored for the backward pass to compute gradients. 
However, their memory usage may vary depending on the training framework. 
In PyTorch, the temporary gradients of intermediate tensor variables can also be viewed as activations.
Compared to other components, buffers consume a relatively small amount of memory. 
We assume that buffers are always stored in the GPU for subsequent analysis.

\textbf{Mixed Precision Training} \cite{mixed-precision-micikevicius2017mixed}. 
The SOTA approach to train large models utilizes both the half-precision floating-point (FP16) format and the single-precision floating-point (FP32) format during training. 
Parameters, gradients, and activations are stored and computed in FP16 to reduce memory usage and improve efficiency. 
Meanwhile, the accumulation operator in the optimizer update is sensitive to underflow in low-precision formats.
The master weight, which is an FP32 copy of the parameters, is used to accumulate gradients in each optimizer update and is rounded to FP16 parameters before the forward pass.
In this case, the memory usage of parameters, gradients, and activations is halved, but the memory usage of optimizer states is increased due to the addition of the master weight.
For example, if we use Adam and the model size is $M$, training requires $2M$ bytes for parameters, $2M$ bytes for gradients, and $12M$ bytes for optimizer states.

After understanding how memory is consumed during training, we discuss how to reduce the memory usage.

\subsection{Reduce the Memory Usage of Model States}

Model states consume a large part of memory when training large models.
ZeRO and memory offloading are two effective techniques used to reduce the memory usage of model states.

\textbf{ZeRO-n}. ZeRO is an optimization technique for DDP.
In traditional DDP, all types of memory usage, except for activations, are replicated across GPUs. 
ZeRO partitions model states to eliminate memory redundancy.
It offers three stages for three degrees of memory partitioning.
ZeRO-1 partitions optimizer states among GPUs, where each compute device only updates a partition of the model during optimizer updates. 
ZeRO-2 partitions gradients and optimizer states, with gradients scattered after their reduce collective communication.
ZeRO-3 partitions parameters, gradients, and optimizer states, with parameters required to be gathered before computing.
Thus, when ZeRO-3 is enabled, training on 4 GPUs only requires $\frac{1}{2}M$ bytes for parameters, $\frac{1}{2}M$ bytes for gradients, and $3M$ bytes for optimizer states on each device.

\textbf{Memory Offloading}. Researchers store model states in CPU memory to address the shortage of GPU memory.
L2L \cite{l2l-pudipeddi2020training} uploads each layer to the GPU before computations and offloads them to the CPU afterward. 
ZeRO-Offload not only utilizes CPU memory resources but also uses CPU compute resources.  
It offloads both optimizer states and optimizer update operators to the CPU. 
In the heterogeneous training paradigm proposed by ZeRO-Offload, FP16 parameters, gradients, and activations are placed in GPU memory, while FP32 optimizer states are placed in CPU memory. 
Before each forward pass, FP16 parameters are updated by the FP32 master weight stored in CPU. 
After each backward pass, FP16 gradients are transferred to CPU memory for CPU optimizer update. 
ZeRO-Infinity combines ZeRO-3 and memory offloading, where all model states are offloaded to CPU memory, and training only requires a buffer to store the currently computed parameters on each device.

\subsection{Reduce the Memory Usage of Activations}

The memory usage of activations and model states is within the same order of magnitude. 
However, the memory usage of activations can be significantly reduced through recomputation.

\textbf{Recomputation}. 
Activation Checkpointing (AC) \cite{ac-chen2016training}, also known as gradient checkpointing, is a technique used to decrease the memory usage of activations. 
AC does not store intermediate variables in the forward pass; it recomputes the forward operators during the backward pass. 
Thus, the number of intermediate variables stored for the backward pass is reduced, thereby reducing the peak memory usage of activations.
For instance, the 4B parameter GPT-2 model trained with a sequence length of 1K and a batch size of 2 requires 14.9GB of memory to store activations. 
Enabling AC reduces the memory usage of activations to 1.3GB.

\section{Elixir: Overview}

\begin{figure}[t]
    \centering
    \label{fig:overview}
    \includegraphics[width=\linewidth]{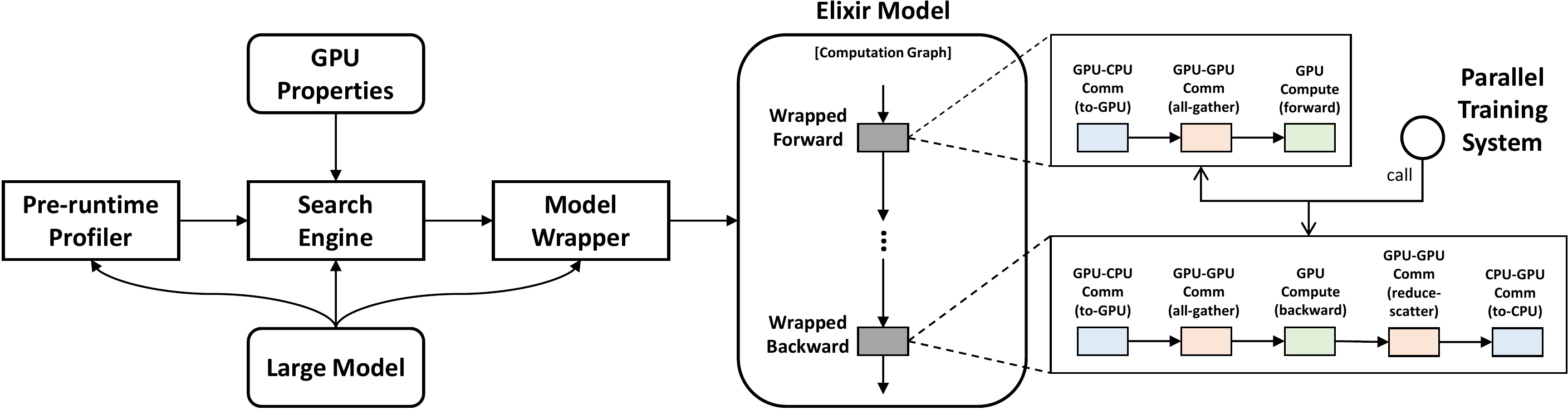}
    \caption{Coarse-grained workflow.}
\end{figure}

We propose Elixir to make hardware-efficient large model training accessible to everyone.
It comprises three main components: a \textbf{pre-runtime profiler}, a \textbf{search engine}, and a \textbf{parallel training system}. 
The coarse-grained workflow of Elixir is quite simple (as shown in Fig.\ref{fig:overview}).
The pre-runtime profiler scans the given large model and generates its training information.
With the knowledge of training details, the search engine identifies the optimal configuration and passes it to the model wrapper, which initializes an Elixir model according to the given configuration.
All compute operators in the Elixir model are transformed into wrapped operators that automatically handle communication through the parallel training system.
Before discussing more details about the components, we need to introduce some foundational concepts first.


\textbf{Chunk}. 
We borrow this concept from PatrickStar \cite{pstar-fang2022parallel}. 
We flatten a group of parameters and concatenate them into a one-dimensional parameter with a fixed length (as shown in Fig.\ref{fig:chunks} left). 
The insight is that one-dimensional parameters of a uniform length are more convenient to manage compared to parameters with different sizes and shapes. 
Moreover, we can use relatively large chunks as the communication units to fully utilize communication bandwidth.

\textbf{rCache}.
rCache is a new memory layer allocated in GPU DRAM. 
We establish that only data in rCache can be duplicated across GPUs, and no other memory layers are permitted to have redundant data.
In our implementation, rCache has $n_{\text{block}}$ storage blocks of the same length to store gathered chunks.
Chunks are partitioned by default in our distributed solution, we gather them into rCache before compute operators (as shown in Fig.\ref{fig:chunks} right).
Since the calling order of chunks can be acquired from the pre-runtime profiler, we apply Belady's algorithm as the replacement policy of rCache.

\subsection{Main Components}

We take a closer look at the three main components mentioned above.

\textbf{Pre-runtime Profiler}.
We use a pre-runtime profiler to profile the training step before running models. 
There are two main reasons why the profiler needs to be pre-runtime.

(1) Running large models requires complicated parallel techniques.
Different types of parallelism can generate various computation graphs. 
Therefore, profiling large models before running them is a more robust way to collect computation graphs.
(2) Dynamically allocating blocks during training leads to a large amount of memory fragmentation.
For long-term active variables such as storage blocks in rCache and optimizer states, we should allocate them before training to avoid memory fragmentation. 
Only short-term variables such as activations should be dynamically allocated during training.

\textbf{Search Engine}.
Based on the profiling results, an optimal configuration will be automatically found.
In our implementation, the search engine needs to specify the length of chunks $C$, the number of blocks allocated in rCache $n_{\text{block}}$, and the number of chunks placed in CPU.
More details about the search engine can be found in Section \ref{sec:optimal}.

\textbf{Parallel Training System}.
The parallel training system is responsible for the memory management during training.
Except activations which are managed by PyTorch, parameters, gradients, and optimizer states are managed by Elixir.
In our implementation, we wrap all compute operators with GPU-GPU and GPU-CPU communication to ensure that parameters are complete before computations and gradients are partitioned after reduction communications.

\section{Elixir: Cost Analysis}

We analyze the memory usage and communication volume of our parallel training strategy, showing that it can achieve comparable results to ZeRO-2 and ZeRO-3 when the configuration is at the boundary points.
Furthermore, we demonstrate how Elixir finds opportunities for optimization.
In our analysis, we define the following variables: $N$ as the number of GPUs, $M$ as the model size, $L_c$ as the precision length in computations, $L_{os}$ as the precision length in optimizer updates, and $F_{os}$ as the additional memory overhead of the optimizer, $C$ as the chunk length, $S=n_{\text{block}}C$ as the aggregate length of chunks.
In practice, the waste rate caused by packaging all the parameters into chunks is less than 4\%, so we assume $S \approx M$.


\begin{figure}[t]
    \centering
    \includegraphics[width=\linewidth]{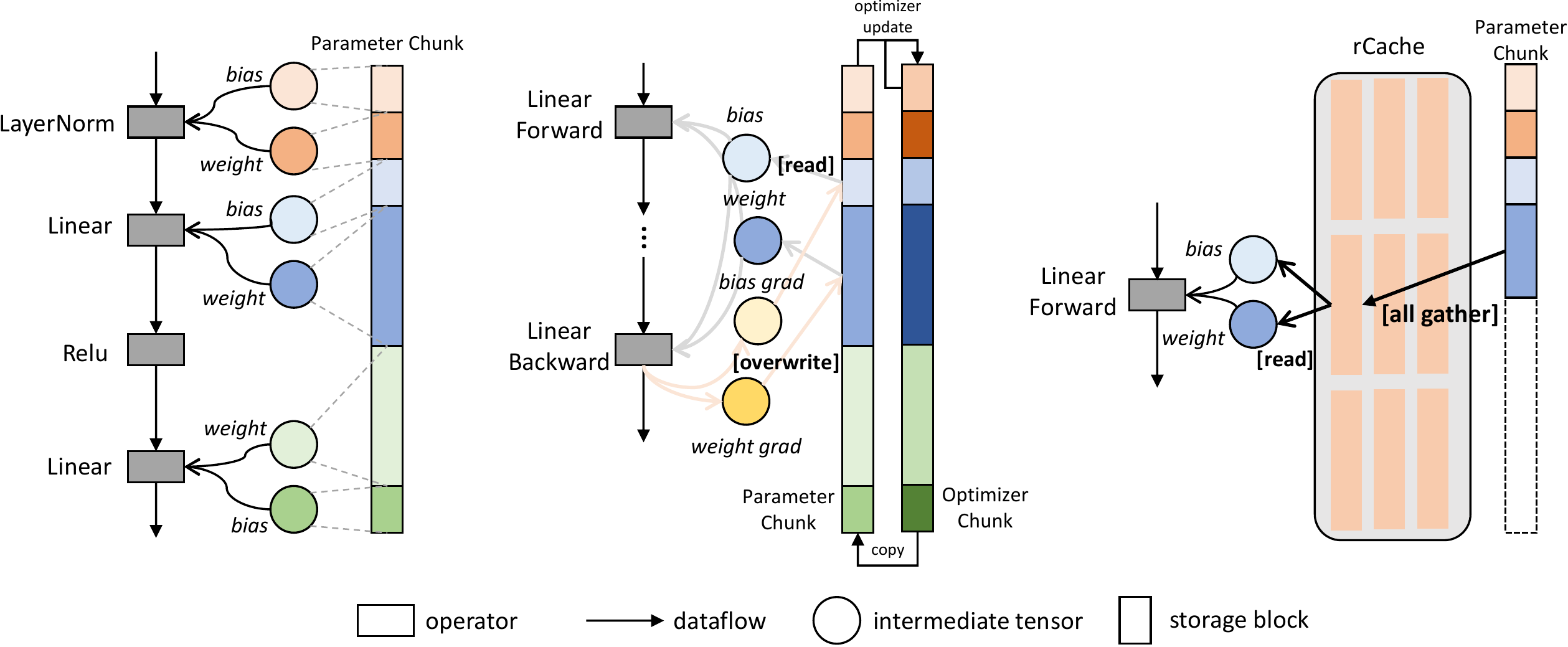}
    \caption{Left: The data of each parameter is compacted into a parameter chunk.
    Center: During the forward and backward pass, each operator reads the data of relevant parameters from their corresponding parameter chunks. 
    Once the gradient calculation of a parameter is finished, it is no longer used in subsequent operators. 
    We use the gradient to overwrite the data in the parameter chunk.
    Right: Chunks should be gathered in rCache before reading their data. 
    After the backward pass, chunks should be scattered and then used to update its distributed shards.}
    \label{fig:chunks}
\end{figure}


\subsection{Total Memory Usage of a Chunk}

We create two chunks for the same group of parameters: the parameter chunk and the optimizer chunk, which are used for computation and optimizer updates, respectively.
We omit gradient chunks because we reuse parameter chunks to store gradients \cite{pstar-fang2022parallel}.
As the gradient of a parameter has the same size as its data, we replace the data of each parameter in the parameter chunk with its generated gradient (as shown in Fig.\ref{fig:chunks} center). 
Once all the parameters within a chunk have completed gradient calculations, that chunk is transformed into a gradient chunk.
We utilize the transformed parameter chunk and optimizer chunk to perform one optimizer update, so each parameter chunk and its paired optimizer chunk must be stored on the same device.
In summary, the total memory usage of a chunk is $\frac{L_cC + L_{os}F_{os}C}{N}$.

\subsection{Compared with ZeRO}

\begin{table}\small
  \renewcommand{\arraystretch}{1.4}
  \caption{Memory usage and communication volume of different parallel distributed strategies. 
  $\epsilon$ is the size of a buffer used to store gathered parameters.}
  \label{cost-comparision}
  \centering
  \begin{tabular}{cccc} 
    \toprule
    Name                    & GPU Memory (per GPU)                          & GPU-CPU Comm          & GPU-GPU Comm \\ 
    \midrule
    DDP                     & $(L_c + L_c + L_{os}F_{os})M$                 & 0                     & $2L_cM$ \\
    ZeRO-1                  & $(L_c + L_c)M + \frac{(L_{os}F_{os})M}{N}$    & 0                     & $2L_cM$      \\
    ZeRO-2 $\rightarrow$ offload        & $L_cM + \frac{(L_c + L_{os}F_{os})M}{N}$ $\rightarrow$ $L_cM$     & 0 $\rightarrow$ $2L_cM$   & $2L_cM$        \\
    ZeRO-3 $\rightarrow$ offload        & $\frac{(L_c + L_c + L_{os}F_{os})M}{N}$ $\rightarrow$ $\epsilon$  & 0 $\rightarrow$ $4L_cM$   & $4L_cM$         \\
    rCache-max $\rightarrow$ offload    & $L_cS + \frac{(L_{c} + L_{os}F_{os})S}{N}$ $\rightarrow$ $L_cS$   & 0 $\rightarrow$ $2L_cS$   & $2L_cS$      \\
    rCache-min $\rightarrow$ offload    & $L_cC + \frac{(L_c  + L_{os}F_{os})S}{N}$ $\rightarrow$ $L_cC$    & 0 $\rightarrow$ $4L_cS$   & $4L_cS$       \\
    \bottomrule
  \end{tabular}
\end{table}

We demonstrate the memory usage and communication volume of various distributed parallel training strategies in Table~\ref{cost-comparision}, where activation checkpointing is enabled. 
The rCache-max strategy is comparable to ZeRO-2.
It sets $n_{\text{blocks}}$ to its maximum number, $n_{\text{chunks}}$, resulting in each chunk only being gathered in the forward pass and reduced in the backward pass. 
The rCache-min strategy is comparable to ZeRO-3. 
The rCache-min strategy limits $n_{\text{blocks}}$ to its minimum number, 1, allowing only one chunk to be gathered during the training step.
In the context of training a large model on a small cluster, we assume that $n_{\text{chunks}} > N$, where we have $C < \frac{M}{N}$. 
When the rCache size is set between the minimum and maximum values, the memory usage is between that of rCache-min and rCache-max, and the communication volume is between that of rCache-max and rCache-min.
In this way, Elixir trades off memory usage and communication volume to utilize free GPU memory.

When the offloading option is enabled, we observe that the GPU-CPU communication volume remains the same as the GPU-CPU communication volume. 
It should be noted that ZeRO-3 Offload and rCache-min Offload do not have the same memory usage. 
In practice, we have found that the latter one utilizes less GPU memory.
Similarly, Elixir can also trade off memory usage and CPU optimizer updates to improve efficiency.

\subsection{Communication Overlap}

We asynchronously prefetch the next-used chunk based on the calling order of chunks from the pre-runtime profiler.
By using streams concurrently, we can overlap the all-gather communication in prefetch operations with the computation of currently active chunks in rCache.  
We assume that communication is not a bottleneck for efficiency and can be perfectly overlapped with computation when training on a small cluster.
Therefore, GPU-GPU communication is not included in the subsequent configuration search analysis.

\section{Elixir: Find the Optimal Configuration} \label{sec:optimal}

The search engine utilizes the profiling results from the pre-runtime profiler to identify an optimal configuration.
The engine obtains the allowed memory usage, denoted as $U_\text{allowed}$, and finds the optimal chunk length $C$ (See more details in Appendix.\ref{apd:algo}).

\subsection{Optimal Offloading Configuration}

We approach finding the optimal configuration as an optimization problem. 
Initially, we set the size of rCache to 1 and offload all chunks to CPU memory.
Our goal is to maximize training throughput within a budget of $U_\text{allowed}$.
There are two options for utilizing the available GPU memory.
The first option is to expand the rCache size to cache more recently used chunks. 
The second option is to upload chunks and their optimizer updates to GPUs. 
We analyze the benefits of both options.

\begin{figure}[t]
    \centering
    \includegraphics[width=0.8\linewidth]{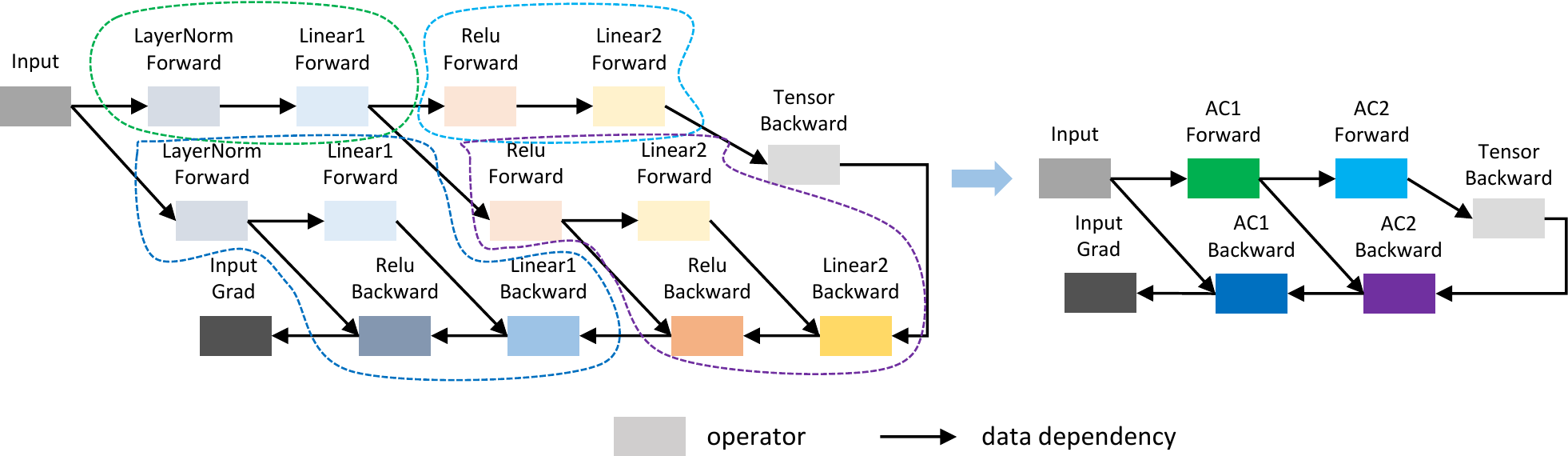}
    \caption{In the left graph, each parameter is called twice in the forward functions and once in the backward functions, resulting in an uncommon computation graph. 
    In the right graph, we consider the AC function as an operator, and each parameter is only called once in the backward pass.}
    \label{fig:graph-ac}
\end{figure}

We introduce the definition of the common computation graph (as shown in Fig.\ref{fig:graph-ac} right).
In most LLM implementations, each parameter is called once during the forward pass and once during the backward pass, except for the weight in the embedding layer. 
The calling order of parameters in the backward pass is exactly the reverse of the calling order in the forward pass.

The computation graph is no longer common when activation checkpointing (AC) is enabled, as each parameter is called twice in the backward pass (as shown in Fig.\ref{fig:graph-ac} left). 
To address this, we consider the computation graph from a coarse-grained perspective, treating an AC function as a compute operator. 
As shown in Fig.\ref{fig:graph-ac}, we transform an uncommon computation graph into a common computation graph.

Our strategy is to extend rCache to ensure that it is large enough to cache all the chunks used in each coarse-grained compute operator (See more details in Appendix.\ref{apd:ac}).
Therefore, we have the following observations: (1) Each chunk is gathered once into rCache in the forward pass and once in the backward pass. (2) In the backward pass, the order of replicated chunks is the exact reverse of the order in the forward pass.

We define the following variables to analyze the benefits: $B_{g2c}(n)$ and $B_{c2g}(n)$ represent the aggregate communication bandwidth of $n$ processes when transferring data from GPU to CPU and from CPU to GPU, respectively. 
$V_g(n)$ and $V_c(n)$ represent the aggregate update velocity with $n$ processes on GPU and CPU, respectively.

\begin{equation}
    I(n) = \frac{1}{L_c} \left( \frac{L_cC}{B_{g2c}(n)} + \frac{L_cC}{B_{c2g}(n)} \right)
\end{equation}

We analyze the time saved when extending a single storage block for rCache.
With one more storage block available, rCache is able to cache one more chunk at the end of the forward pass. 
This cached chunk does not require extra offloading communication in the backward pass. 
The normalized benefit is denoted as $I(n)$.

\begin{equation}
    J(n) = \frac{n}{L_c + L_{os}F_{os}} \left[ \left( \frac{L_{os}C}{B_{c2g}(n)} + L_c I(n) + \frac{L_cC}{B_{g2c}(n)} \right) + \left( \frac{C}{V_c(n)} - \frac{C}{V_g(n)} \right) \right]
\end{equation}

We analyze the time saved when uploading a chunk to the GPU. 
We eliminate offloading communication for this chunk and replace its CPU optimizer update with a GPU optimizer update. 
The normalized benefit is denoted as $J(n)$.

We can compare $I(n)$ and $J(n)$ to determine whether to prioritize uploading chunks or expanding rCache.
For instance, if $J(n) > I(n)$, we should prioritize uploading as many chunks as possible.
If there is still available space, we can then shift our focus to extending rCache.

\section{Experiments}

We evaluate the performance of using Elixir to train GPT models on various training configurations\footnote[1]{The benchmark code is available at \url{https://github.com/hpcaitech/Elixir}}.
We observe that Elixir runs faster than existing SOTA solutions: up to 3.4$\times$ speedup on A100 80GB, and 3$\times$ speedup on A100 40GB (see Appendix.\ref{apd:exp_res}).

\subsection{Experiments Settings}

We conducted our experiments using two types of A100 compute nodes. 
One type is equipped with 4 A100 80GB PCIe GPUs and 400GiB CPU DRAM. 
The other type has 4 A100 40GB SXM GPUs, NVLink \cite{nvlink}, and 500GiB CPU DRAM.  
See Appendix.\ref{apd:hardware} for more details about the hardware performance.

We exclusively evaluate GPT-2 models, as large language models typically share the same transformer \cite{attention-vaswani2017attention} architecture as GPT-2. 
We vary the model size, batch size, number of GPUs, and GPU capacity to assess how the performance changes across different training configurations.
Activation checkpointing and mixed-precision training are enabled in our experiments. 
To measure training efficiency, we utilize TFLOPS as the metric, which is defined as $8MD$ \cite{nd-kaplan2020scaling}, where $M$ represents the model size and $D$ represents the number of tokens.

Our baselines include Megatron-LM, FSDP \cite{fsdp-zhao2023pytorch}, and DeepSpeed. 
We do not include PP because it is commonly employed across compute nodes \cite{3d-parallel-narayanan2021efficient}.
The training efficiency of DeepSpeed is measured as the highest throughput selected from ZeRO-2, ZeRO-2 Offload, ZeRO-3, and ZeRO-3 Offload. 
Notice that GPT-2 models evaluated with Elixir, FSDP, and DeepSpeed are sourced from HuggingFace \cite{hf-wolf2020transformers}, while GPT-2 models evaluated with Megatron are implemented by itself.
We also apply FlashAttention \cite{fa-dao2022flashattention} in GPT-2 models.

\subsection{Experiments Results}

Our experiments focus on validating the robustness of Elixir.
We vary the training arguments to modify the memory usage settings, including using different model sizes to adjust the memory usage of model states, varying GPU capacities and the number of GPUs to modify the aggregate GPU memory size, and utilizing various batch sizes to alter the memory usage of activations. 
The complete results can be found in Appendix.\ref{apd:exp_res}. 
Here, we have selected a representative subset of the data for analysis.

\begin{table}[htb]\scriptsize
  \caption{The profiled training efficiency (TFLOPS) when the batch size per GPU is set to 8.}
  \label{exp:model_size}
  \centering
  \begin{tabular}{c | c | c c c c | c c c c c} 
    \toprule
     & & \multicolumn{4}{c|}{A100 40GB SXM} & \multicolumn{4}{c}{A100 80GB PCIe} \\
    Number of GPUs      & Model     & Elixir    & Megatron      & FSDP      & DeepSpeed & Elixir    & Megatron      & FSDP      & DeepSpeed \\
    \midrule
    \multirow{4}{*}{1} 
                    & 4b   & \textbf{82}        & OOM           & OOM       & 46        & 147       & \textbf{155}           & 148       & 140 \\
                    & 10b  & \textbf{66}        & OOM           & OOM       & 46        & \textbf{84}        & OOM           & OOM       & 58  \\
                    & 15b  & \textbf{77}        & OOM           & OOM       & 37        & \textbf{77}        & OOM           & OOM       & 62  \\
                    & 20b  & \textbf{69}        & OOM           & OOM       & 36        & \textbf{73}        & OOM           & OOM       & 30  \\
    \midrule
    \multirow{4}{*}{2} 
                    & 4b   & \textbf{152}       & OOM           & OOM       & 125       & \textbf{150}       & 148           & 144       & 144 \\
                    & 10b  & \textbf{91}        & OOM           & OOM       & 55        & \textbf{145}       & OOM           & OOM       & 72  \\
                    & 15b  & \textbf{96}        & OOM           & OOM       & 55        & \textbf{104}       & OOM           & OOM       & 83  \\
                    & 20b  & \textbf{89}        & OOM           & OOM       & 46        & \textbf{101}       & OOM           & OOM       & 48  \\
    \midrule
    \multirow{4}{*}{4} 
                    & 4b   & \textbf{157}       & 129           & 131       & 153       & \textbf{137}       & 97            & 130       & 131 \\
                    & 10b  & \textbf{138}       & OOM           & OOM       & 60        & \textbf{145}       & 116           & OOM       & 130  \\
                    & 15b  & \textbf{129}       & OOM           & OOM       & 62        & \textbf{150}       & 138           & OOM       & 148  \\
                    & 20b  & \textbf{123}       & OOM           & OOM       & 60        & \textbf{139}       & OOM           & OOM       & OOM  \\
              
    \bottomrule
  \end{tabular}
\end{table}

\textbf{Increasing the Model Size}.
The results are presented in Table~\ref{exp:model_size} which demonstrates the impact of increasing the memory usage of model states.
Megatorn and FSDP encounter OOM errors in most cases due to their requirement for a large aggregate GPU memory space.
DeepSpeed can consistently train large models, partly because it is measured across four distributed configurations.
Elixir not only avoids OOM errors during training but also achieves the highest throughput.
On A100 40GB, Elixir achieves 2$\times$ speedup in most cases.
On A100 80GB, Elixir also achieves up to 2$\times$ speedup.

We observe that Elixir achieves less speedup on small models or GPUs with more memory space, as the aggregate GPU memory is sufficient when training small models or with an abundance of GPUs. 
In this case, memory offloading is not necessary, and current SOTA solutions have nearly reached optimal efficiency.

We also observe that the training efficiency drops when increasing the model size, because memory offloading brings extra communication between the CPU and GPUs.
Elixir is able to alleviate this negative impact when there are more GPUs.
While DeepSpeed has significant performance drop when increasing the model size due to the change of the used distributed solution.
This observation proves that Elixir utilizes GPU memory more intelligently.

\begin{table}[htb]\scriptsize
  \caption{The profiled training efficiency (TFLOPS) when the number of GPUs is set to 4. 
  Due to the lack of CPU memory on the A100 80GB compute node, DeepSpeed occurs OOM error when training the 20b GPT-2 model.}
  \label{exp:batch_size}
  \centering
  \begin{tabular}{c | c | c c c c | c c c c c} 
    \toprule
     & & \multicolumn{4}{c|}{A100 40GB SXM} & \multicolumn{4}{c}{A100 80GB PCIe} \\
    model      & batch size    & Elixir    & Megatron      & FSDP      & DeepSpeed & Elixir    & Megatron      & FSDP      & DeepSpeed \\
    \midrule
    \multirow{4}{*}{4b} 
                    & 4     & 144           & 123           & 131       & 139   & 114       & 79        & 106       & 109 \\
                    & 12    & 164           & 107           & OOM       & 153   & 147       & 99        & 141       & 140  \\
                    & 16    & 167           & OOM           & OOM       & 164   & 154       & 99        & 146       & 147  \\
    \midrule
    \multirow{4}{*}{10b} 
                    & 4     & 112           & OOM           & OOM       & 36    & 124       & 112       & OOM       & 108 \\
                    & 12    & 136           & OOM           & OOM       & 77    & 153       & 113       & OOM       & 146  \\
                    & 16    & 151           & OOM           & OOM       & 91    & 164       & 105       & OOM       & 158  \\
    \midrule
    \multirow{4}{*}{15b} 
                    & 4     & 97            & OOM           & OOM       & 34    & 126       & 142        & OOM       & 117 \\
                    & 12    & 144           & OOM           & OOM       & 82    & 167       & 138        & OOM       & 163  \\
                    & 16    & 150           & OOM           & OOM       & 102   & 175       & OOM        & OOM       & 164  \\
    \midrule
    \multirow{4}{*}{20b} 
                    & 4     & 91            & OOM           & OOM       & 33    & 102       & OOM        & OOM       & OOM \\
                    & 12    & 141           & OOM           & OOM       & 79    & 151       & OOM        & OOM       & OOM  \\
                    & 16    & 146           & OOM           & OOM       & 95    & 161       & OOM        & OOM       & OOM  \\
    \bottomrule
  \end{tabular}
\end{table}

\textbf{Increasing the Batch Size}.
We increase the batch size to increase the memory usage of activations, and the profiled results is shown in Table~\ref{exp:batch_size}.
For Elixir, FSDP, and DeepSpeed, their training throughput become higher when increasing the batch size.
Meanwhile, the training throughput of Megatron keeps the same.
The reason is that the communication volume of the previous three solutions is the model size but the communication volume of Megatron increases linearly with the batch size.

We observe that the speedup ratio of Elixir to DeepSpeed becomes smaller when increasing the batch size.
There are two aspects to explain this phenomenon. 
Increasing the batch size increases the memory usage of activations and the time elapsed on the computation.
Thus, there is no enough available GPU memory for Elixir to arrange.
Also, the proportion of computation time increases, making the optimization in communications and CPU computations less effective.
Despite these limitations, Elixir still achieves a 1.5$\times$ speedup on A100 40GB.

\section{Limitations}

We focus on small GPU clusters, and our research does not involve large-scale model training.
Current research suggests that using the complicated combined parallelism plans is a better choice than relying only one distributed solution when training on a large network with intricate communication topology \cite{3d-parallel-narayanan2021efficient, alpa-zheng2022alpa}.

\section{Conclusion}

Large model training becomes a fundamental approach widely applied in DL domains.
Memory partitioning and memory offloading are necessary techniques to train large models on limited compute devices.
Thus, automating the offloading training is the key to truly make the large model training accessible to everyone.
To the best of our knowledge, we are the first to consider both partitioning and offloading together and achieve the optimal configuration within our search space.
Our findings suggest that using a dynamic distributed strategy based on profiling results can result in tripled throughput compared to a rigid distributed strategy.
We hope that our work can assist the majority of AI researchers.

\bibliographystyle{unsrt}
\bibliography{main}


\newpage

\appendix

\section{Algorithm Details} \label{apd:algo}

\subsection{Allowed Memory Usage}
The search engine depends on the allowed memory usage, denoted as $U_\text{allowed}$.

\begin{equation*}
    U_\text{allowed} = F_{\text{alloc}} \cdot \left(\text{capacity}_\text{gpu} - U_{\text{buffer}} - F_\text{frag} \cdot U_\text{activation} \right)
\end{equation*}

In the equation, $U_{\text{buffer}}$ and $U_\text{activation}$ represent the memory usage of buffers and activations, respectively. 
We receive these two variables from the pre-runtime profiler.
Due to the impact of generated memory fragments during training, we must reserve additional GPU memory for activations. 
To account for this reservation, we multiply the memory usage of activations by a ratio greater than 1, denoted as $F_{\text{frag}}$. 
The value of $F_{\text{alloc}}$ is used to estimate the proportion of GPU memory that Elixir can effectively use. 
By default, $F_{\text{alloc}}$ is set to 0.95, and $F_\text{frag}$ is set to 1.25.

\subsection{Grouping Method and Optimal Chunk Size}

For a given chunk size, we sort all parameters into a sequence based on their order of use during the forward pass, assuming that each parameter is used only once. 
Parameters used multiple times, such as the weight of the embedding layer, are rare in language models, and we handle them by using ZeRO-2. 
We iterate through this sequence, maintaining a current chunk and attempting to group each parameter with its preceding parameter. 
If the current chunk is insufficient for the parameter, we close it and assign a new chunk.

We define the optimal chunk size as the size that minimizes the total number of bytes replaced in rCache during one training step.
To determine this size, we simulate various chunk sizes, calculate the number of replacements for each, and select the best value. 
To search more values within a given time, we implement the simulation algorithm in C++ for better efficiency.


\subsection{Activation Checkpointing Detector} \label{apd:ac}

We detect the parameters accessed by each activation checkpointing (AC) function using the pre-runtime profiler.
We define the buffer size of AC as the maximum summation of the number of elements in all parameters accessed by each AC function.
The first priority of Elixir is to ensure that the rCache size is greater than the buffer size of AC. 
If $U_\text{allowed}$ is smaller than the buffer size of AC, we extend rCache as much as possible and keep all chunks in CPU memory.

\section{Experiments Details}

\subsection{Hardware Performance} \label{apd:hardware}

We profiled the hardware performance on our development server and AWS cloud server. 
The results are shown in Table~\ref{hp:dev} and Table~\ref{hp:aws}. 
Additionally, the hardware performance on NSCC is similar to our development server, with the exception that the 4 GPUs are linked by high-bandwidth NVLINK, reaching 200GB/s.
Comparing the two tables, we observe that the communication bandwidth between the GPU and CPU on AWS is significantly lower.
When comparing the two tables, we observe a difference in the communication bandwidth between the GPU and CPU on AWS, which is notably lower. 
Consequently, memory offloading leads to a performance drop on AWS.

\begin{table}[htb]
  \caption{Hardware performance (GB/s) profiled on our development server.}
  \label{hp:dev}
  \centering
  \begin{tabular}{cccccc} 
    \toprule
    \multicolumn{6}{c}{Development Server} \\
    $n$ & $B_{g2g}(n)$ & $B_{c2g}(n)$  & $B_{g2c}(n)$ & $V_g(n)$ & $V_c(n)$ \\
    \midrule
    1   &   -       & 22        & 16    & 50    & 5    \\
    2   &   201     & 50        & 40    & 100   & 6.5  \\
    4   &   58      & 70        & 60    & 200   & 7.5  \\
    \bottomrule
  \end{tabular}
\end{table}

\begin{table}[htb]
  \caption{Hardware performance (GB/s) profiled on an p4d.24xlarge instance provided by AWS.}
  \label{hp:aws}
  \centering
  \begin{tabular}{cccccc} 
    \toprule
    \multicolumn{6}{c}{AWS Cloud Server} \\
    $n$ & $B_{g2g}(n)$ & $B_{c2g}(n)$  & $B_{g2c}(n)$ & $V_g(n)$ & $V_c(n)$ \\
    \midrule
    1   &   -       & 12        & 13    & 44    & 3.7    \\
    2   &   188     & 12        & 13    & 86    & 5.6  \\
    4   &   214     & 25        & 26    & 171   & 5  \\
    \bottomrule
  \end{tabular}
\end{table}

\subsection{Model Configurations}

GPT-2 model configurations are shown in Table~\ref{mc:gpt}.

\begin{table}[h]
  \caption{GPT-2 model configurations.}
  \label{mc:gpt}
  \centering
  \begin{tabular}{cccc} 
    \toprule
    name & hidden size & num layer & num attention heads \\
    \midrule
    GPT2-4b     & 3072  & 32    & 24 \\
    GPT2-10b    & 4096  & 48    & 32  \\
    GPT2-15b    & 8192  & 18    & 64 \\
    GPT2-20b    & 8192  & 24    & 64 \\
    \bottomrule
  \end{tabular}
\end{table}

\section{Full Experimental Results} \label{apd:exp_res}

The complete experimental results are presented in Table~\ref{aer:40gb} and Table~\ref{aer:80gb}. 
Elixir achieves a speedup of up to 3x on A100 40GB and 3.4x on A100 80GB.

\begin{table}[htb]\footnotesize
  \caption{Experimental results (TFLOPS) on A100 40GB.}
  \label{aer:40gb}
  \centering
  \begin{tabular}{c | c | c | c c c c c c c | c}
    \toprule
    $M$ & $n$ & $bs$ & Elixir & Megatron & FSDP & DS-Z2 & DS-Z3 & DS-Z2-o & DS-Z3-o & speedup \\
    \hline
    \multirow{12}{*}{4}
      & \multirow{4}{*}{1}
        & 4 & 62.0 & OOM & OOM & OOM & OOM & 26.4 & 19.4 & 2.35x\\
        & & 8 & 82.7 & OOM & OOM & OOM & OOM & 46.3 & 40.5 & 1.79x\\
        & & 12 & 102.5 & OOM & OOM & OOM & OOM & 64.3 & 50.7 & 1.59x\\
        & & 16 & 113.3 & OOM & OOM & OOM & OOM & 77.6 & 60.6 & 1.46x\\
    \cline{2-11}
      & \multirow{4}{*}{2}
        & 4 & 137.3 & 138.0 & OOM & OOM & 102.0 & 33.8 & 25.7 & 0.99x\\
        & & 8 & 152.2 & OOM & OOM & OOM & 125.7 & 53.3 & 46.0 & 1.21x\\
        & & 12 & 148.5 & OOM & OOM & OOM & OOM & 73.6 & 64.3 & 2.02x\\
        & & 16 & 145.8 & OOM & OOM & OOM & OOM & 81.0 & 75.4 & 1.8x\\
    \cline{2-11}
      & \multirow{4}{*}{4}
        & 4 & 144.6 & 123.8 & 131.2 & 137.1 & 139.9 & 34.1 & 28.9 & 1.03x\\
        & & 8 & 157.7 & 130.0 & 131.3 & 153.2 & 152.4 & 60.7 & 54.8 & 1.03x\\
        & & 12 & 164.3 & 107.1 & OOM & 161.4 & 160.1 & 77.4 & 69.4 & 1.02x\\
        & & 16 & 167.1 & OOM & OOM & 164.9 & 162.5 & 91.4 & 89.2 & 1.01x\\
    \hline
    \multirow{12}{*}{10}
      & \multirow{4}{*}{1}
        & 4 & 47.9 & OOM & OOM & OOM & OOM & 27.6 & 21.1 & 1.73x\\
        & & 8 & 66.5 & OOM & OOM & OOM & OOM & 46.5 & 39.1 & 1.43x\\
        & & 12 & 88.0 & OOM & OOM & OOM & OOM & 60.7 & 54.8 & 1.45x\\
        & & 16 & 98.7 & OOM & OOM & OOM & OOM & 80.7 & 65.9 & 1.22x\\
    \cline{2-11}
      & \multirow{4}{*}{2}
        & 4 & 64.6 & OOM & OOM & OOM & OOM & 30.3 & 26.3 & 2.14x\\
        & & 8 & 91.2 & OOM & OOM & OOM & OOM & 55.9 & 49.0 & 1.63x\\
        & & 12 & 103.8 & OOM & OOM & OOM & OOM & 70.6 & 62.8 & 1.47x\\
        & & 16 & 118.3 & OOM & OOM & OOM & OOM & 80.4 & 75.6 & 1.47x\\
    \cline{2-11}
      & \multirow{4}{*}{4}
        & 4 & 112.6 & OOM & OOM & OOM & OOM & 36.5 & 31.9 & \textbf{3.09x}\\
        & & 8 & 138.4 & OOM & OOM & OOM & OOM & 60.3 & 55.3 & 2.3x\\
        & & 12 & 136.3 & OOM & OOM & OOM & OOM & 78.0 & 75.2 & 1.75x\\
        & & 16 & 151.7 & OOM & OOM & OOM & OOM & 92.0 & 88.8 & 1.65x\\
    \hline
    \multirow{12}{*}{15}
      & \multirow{4}{*}{1}
        & 4 & 42.0 & OOM & OOM & OOM & OOM & OOM & 19.2 & 2.19x\\
        & & 8 & 77.6 & OOM & OOM & OOM & OOM & OOM & 37.1 & 2.09x\\
        & & 12 & 93.6 & OOM & OOM & OOM & OOM & OOM & 63.4 & 1.48x\\
        & & 16 & 107.8 & OOM & OOM & OOM & OOM & OOM & 64.6 & 1.67x\\
    \cline{2-11}
      & \multirow{4}{*}{2}
        & 4 & 59.7 & OOM & OOM & OOM & OOM & OOM & 27.0 & 2.21x\\
        & & 8 & 96.4 & OOM & OOM & OOM & OOM & OOM & 55.7 & 1.73x\\
        & & 12 & 109.7 & OOM & OOM & OOM & OOM & OOM & 69.6 & 1.58x\\
        & & 16 & 128.5 & OOM & OOM & OOM & OOM & OOM & 87.3 & 1.47x\\
    \cline{2-11}
      & \multirow{4}{*}{4}
        & 4 & 97.7 & OOM & OOM & OOM & OOM & OOM & 35.0 & 2.79x\\
        & & 8 & 129.1 & OOM & OOM & OOM & OOM & OOM & 62.8 & 2.05x\\
        & & 12 & 144.3 & OOM & OOM & OOM & OOM & OOM & 82.2 & 1.76x\\
        & & 16 & 151.0 & OOM & OOM & OOM & OOM & OOM & 102.5 & 1.47x\\
    \hline
    \multirow{12}{*}{20}
      & \multirow{4}{*}{1}
        & 4 & 42.0 & OOM & OOM & OOM & OOM & OOM & 18.9 & 2.22x\\
        & & 8 & 69.4 & OOM & OOM & OOM & OOM & OOM & 36.6 & 1.9x\\
        & & 12 & 86.2 & OOM & OOM & OOM & OOM & OOM & 48.8 & 1.77x\\
        & & 16 & 108.4 & OOM & OOM & OOM & OOM & OOM & 63.2 & 1.71x\\
    \cline{2-11}
      & \multirow{4}{*}{2}
        & 4 & 55.5 & OOM & OOM & OOM & OOM & OOM & 25.0 & 2.22x\\
        & & 8 & 89.0 & OOM & OOM & OOM & OOM & OOM & 47.0 & 1.89x\\
        & & 12 & 113.6 & OOM & OOM & OOM & OOM & OOM & 64.5 & 1.76x\\
        & & 16 & 130.5 & OOM & OOM & OOM & OOM & OOM & 78.1 & 1.67x\\
    \cline{2-11}
      & \multirow{4}{*}{4}
        & 4 & 91.8 & OOM & OOM & OOM & OOM & OOM & 33.6 & 2.73x\\
        & & 8 & 124.0 & OOM & OOM & OOM & OOM & OOM & 60.2 & 2.06x\\
        & & 12 & 141.2 & OOM & OOM & OOM & OOM & OOM & 79.4 & 1.78x\\
        & & 16 & 146.5 & OOM & OOM & OOM & OOM & OOM & 95.1 & 1.54x\\

    \bottomrule
  \end{tabular}
\end{table}

\begin{table}[htb]\footnotesize
  \caption{Experimental results (TFLOPS) on A100 80GB.}
  \label{aer:80gb}
  \centering
  \begin{tabular}{c | c | c | c c c c c c c | c}
    \toprule
    $M$ & $n$ & $bs$ & Elixir & Megatron & FSDP & DS-Z2 & DS-Z3 & DS-Z2-o & DS-Z3-o & speedup \\
    \hline
    \multirow{12}{*}{4}
      & \multirow{4}{*}{1}
        & 4 & 131.7 & 140.5 & 135.3 & OOM & 122.6 & 32.2 & 16.0 & 0.94x\\
        & & 8 & 147.1 & 155.8 & 148.3 & OOM & 140.3 & 54.1 & 28.6 & 0.94x\\
        & & 12 & 155.0 & 166.6 & 155.6 & OOM & 150.0 & 86.1 & 45.5 & 0.93x\\
        & & 16 & 158.4 & 167.7 & 158.3 & OOM & 155.3 & 82.4 & 53.9 & 0.94x\\
    \cline{2-11}
      & \multirow{4}{*}{2}
        & 4 & 136.5 & 137.6 & 130.1 & 126.8 & 129.1 & 35.3 & 27.3 & 0.99x\\
        & & 8 & 150.6 & 149.0 & 144.5 & 144.0 & 143.7 & 72.5 & 41.1 & 1.01x\\
        & & 12 & 157.4 & 153.1 & 152.6 & 152.8 & 151.5 & 84.8 & 59.8 & 1.03x\\
        & & 16 & 161.4 & 152.5 & 155.6 & 156.5 & 154.5 & 89.7 & 69.7 & 1.03x\\
    \cline{2-11}
      & \multirow{4}{*}{4}
        & 4 & 114.4 & 79.0 & 106.5 & 92.0 & 109.4 & 38.3 & 27.7 & 1.05x\\
        & & 8 & 137.4 & 98.0 & 131.0 & 119.2 & 131.6 & 59.5 & 52.2 & 1.04x\\
        & & 12 & 147.9 & 99.2 & 141.3 & 133.4 & 140.1 & 78.7 & 67.2 & 1.05x\\
        & & 16 & 155.0 & 99.2 & 146.3 & 141.2 & 147.5 & 89.0 & 90.6 & 1.05x\\
    \hline
    \multirow{12}{*}{10}
      & \multirow{4}{*}{1}
        & 4 & 56.0 & OOM & OOM & OOM & OOM & 33.9 & 14.1 & 1.65x\\
        & & 8 & 84.8 & OOM & OOM & OOM & OOM & 58.6 & 29.1 & 1.45x\\
        & & 12 & 102.1 & OOM & OOM & OOM & OOM & 73.1 & 43.9 & 1.4x\\
        & & 16 & 110.3 & OOM & OOM & OOM & OOM & 88.7 & 55.0 & 1.24x\\
    \cline{2-11}
      & \multirow{4}{*}{2}
        & 4 & 130.3 & OOM & OOM & OOM & OOM & 38.4 & 26.1 & \textbf{3.39x}\\
        & & 8 & 145.6 & OOM & OOM & OOM & OOM & 72.8 & 46.7 & 2.0x\\
        & & 12 & 144.4 & OOM & OOM & OOM & OOM & 88.8 & 62.8 & 1.63x\\
        & & 16 & 154.9 & OOM & OOM & OOM & OOM & 108.0 & 66.5 & 1.43x\\
    \cline{2-11}
      & \multirow{4}{*}{4}
        & 4 & 124.8 & 112.8 & OOM & OOM & 108.8 & 40.8 & 29.4 & 1.11x\\
        & & 8 & 145.5 & 116.1 & OOM & OOM & 140.4 & 65.0 & 57.5 & 1.04x\\
        & & 12 & 153.2 & 113.7 & OOM & OOM & 146.2 & 86.4 & 71.9 & 1.05x\\
        & & 16 & 164.3 & 105.8 & OOM & OOM & 158.2 & 89.4 & 87.0 & 1.04x\\
    \hline
    \multirow{12}{*}{15}
      & \multirow{4}{*}{1}
        & 4 & 47.6 & OOM & OOM & OOM & OOM & 36.7 & 18.9 & 1.3x\\
        & & 8 & 77.3 & OOM & OOM & OOM & OOM & 62.3 & 30.1 & 1.24x\\
        & & 12 & 94.2 & OOM & OOM & OOM & OOM & 82.1 & 41.1 & 1.15x\\
        & & 16 & 108.9 & OOM & OOM & OOM & OOM & 98.6 & 57.2 & 1.1x\\
    \cline{2-11}
      & \multirow{4}{*}{2}
        & 4 & 90.0 & OOM & OOM & OOM & OOM & 40.3 & 29.8 & 2.23x\\
        & & 8 & 104.3 & OOM & OOM & OOM & OOM & 83.0 & 47.5 & 1.26x\\
        & & 12 & 133.5 & OOM & OOM & OOM & OOM & 87.0 & 72.2 & 1.53x\\
        & & 16 & 139.9 & OOM & OOM & OOM & OOM & 116.6 & 79.8 & 1.2x\\
    \cline{2-11}
      & \multirow{4}{*}{4}
        & 4 & 126.4 & 142.3 & OOM & OOM & 117.4 & 41.3 & 33.1 & 0.89x\\
        & & 8 & 150.4 & 138.2 & OOM & OOM & 148.5 & 66.1 & 60.5 & 1.01x\\
        & & 12 & 167.1 & OOM & OOM & OOM & 163.9 & 84.1 & 75.0 & 1.02x\\
        & & 16 & 175.7 & OOM & OOM & OOM & 164.3 & 102.4 & 93.3 & 1.07x\\
    \hline
    \multirow{12}{*}{20}
      & \multirow{4}{*}{1}
        & 4 & 48.0 & OOM & OOM & OOM & OOM & OOM & 15.5 & 3.1x\\
        & & 8 & 73.0 & OOM & OOM & OOM & OOM & OOM & 30.4 & 2.4x\\
        & & 12 & 95.8 & OOM & OOM & OOM & OOM & OOM & 42.8 & 2.24x\\
        & & 16 & 105.7 & OOM & OOM & OOM & OOM & OOM & 61.2 & 1.73x\\
    \cline{2-11}
      & \multirow{4}{*}{2}
        & 4 & 63.3 & OOM & OOM & OOM & OOM & OOM & 25.3 & 2.5x\\
        & & 8 & 101.5 & OOM & OOM & OOM & OOM & OOM & 48.4 & 2.1x\\
        & & 12 & 117.4 & OOM & OOM & OOM & OOM & OOM & 61.9 & 1.9x\\
        & & 16 & 129.6 & OOM & OOM & OOM & OOM & OOM & 78.2 & 1.66x\\
    \cline{2-11}
      & \multirow{4}{*}{4}
        & 4 & 103.0 & OOM & OOM & OOM & OOM & OOM & OOM & - \\
        & & 8 & 139.6 & OOM & OOM & OOM & OOM & OOM & OOM & - \\
        & & 12 & 151.3 & OOM & OOM & OOM & OOM & OOM & OOM & - \\
        & & 16 & 161.7 & OOM & OOM & OOM & OOM & OOM & OOM & - \\

    \bottomrule
  \end{tabular}
\end{table}

\end{document}